# Coordinated Dispatch of Energy Storage Systems in the Active Distribution Network: A Complementary Reinforcement Learning and Optimization Approach

Bohan Zhang, Zhongkai Yi, *Member, IEEE*, Ying Xu, *Senior Member, IEEE*, Zhenghong Tu, *Member, IEEE*

*Abstract*—The complexity and nonlinearity of active distribution network (ADN), coupled with the fast-changing renewable energy (RE), necessitate advanced real-time and safe dispatch approach. This paper proposes a complementary reinforcement learning (RL) and optimization approach, namely SA2CO, to address the coordinated dispatch of the energy storage systems (ESSs) in the ADN. The proposed approach leverages RL's capability to make fast decision and address the model inaccuracies, while optimization methods ensure the ADN security. Furthermore, a hybrid data-driven and expert-experience auxiliary neural network is formulated as a rapid security assessment component in the SA2CO algorithm, enabling dynamic switching between RL and optimization methodologies. Simulation results demonstrate the proposed method's effectiveness and scalability in achieving real-time, safe, and economical dispatch of multiple ESSs in the ADN, surpassing the performance of the state-of-the-art RL and optimization methods.

## Nomenclature

**Abbreviations**
AC         Alternating current.
ADN      Active distribution network.
DRL       Deep reinforcement learning.
ESS       Energy storage system.
MDP      Markov decision process.
RE         Renewable energy.
SAC       Soft actor-critic.
SoE        State of energy.

## I. Introduction

### A. Background and Motivation

Deploying distributed renewable energy generation within the distribution network (DN) is an effective strategy for advancing the utilization of sustainable energy [1]. This makes the power flow bidirectional in the DN, forming an ADN [2]. In addition, the configuration of ESSs in the ADN is conducive to promoting the consumption of RE, further balancing the power supply and demand relationship, and reducing the operating cost of the system [3].

The authors are with the School of Electrical Engineering and Automation, Harbin Institute of Technology, Harbin, 150001, P. R. China.

The coordinated dispatch problem of ESSs in the ADN presents several challenges: Firstly, ESSs exhibit nonlinear behavior, complexity, and possess model parameters that are often imprecise [4]. Secondly, ESSs are utilized alongside RE, necessitating that their charging and discharging controls and dispatch strategies be responsive to the rapidly fluctuating characteristics of RE. Lastly, ensuring the safe and stable operation of the ADN is crucial due to its significant impact on public welfare.

Deep reinforcement learning (DRL) effectively addresses the challenges posed by the non-convex nature of model solutions and the inaccuracies in model parameters, and the trained DRL can be employed for real-time dispatch. Nevertheless, the conventional DRL falls short of fulfilling the security requirements for the ADN dispatch. These aforementioned challenges and obstacles necessitate the adoption of a safe RL algorithm to address the coordinated dispatch problem of ESSs within the ADN.

In view of this, this study proposed an optimization and RL complementary decision pattern, and introduced the security assessment architecture into the RL structure. The proposed approach aims to leverages the RL's advantage in making fast decisions and the optimization method's advantage in promoting the system security, which helps to make real-time dispatch decisions for the ESSs in the ADN in an efficient and safe manner.

### B. Related Literature

The existing dispatch problems of ADN in the state-of-the-art literature can be divided into two categories, namely deterministic problems and stochastic problems [5].

If the inaccuracy and uncertainty of model parameters are ignored or simplified, the ADN scheduling problems are essentially deterministic problems. For deterministic problems, the solution methods can be divided into three categories. The first is numerical methods, such as second-order cone convex relaxation, mixed-integer linear programming (MILP) [6], [7]. Although MILP can effectively deal with large and complex optimization problems, it is difficult to deal with high-order nonlinear models [8]. The second is heuristic algorithms. Typical heuristic algorithms applied to DN dispatch problems include genetic algorithms [9], evolutionary algorithms [10], and simulated annealing algorithms. However, heuristic algorithms require a lot of computing time and cannot be applied to real-time scheduling problems. The third is the model predictive control (MPC) method. MPC has the ability to respond quickly and adapt to various constraints, it can

42minimize the operating cost of DN while meeting security constraints [11]. However, MPC has high maintenance costs and a poor ability to deal with model parameter inaccuracies.

From the mathematical point of view, the dispatch problem of ESSs in the ADN is a sequential decision problem with uncertain parameters and intertemporal constraints. For stochastic problems, the solution methods can be divided into two categories. The first category is mathematical optimization methods, such as robust optimization and approximate chance constraints methods [12], [13], [14]. The solutions obtained by mathematical optimization methods to solve stochastic problems are often conservative and take a long time to solve.

The second category is DRL. Proximal Policy Optimization (PPO) is a typical DRL algorithm with an actor-critic architecture [15], which is used to solve the OPF problem of ADN involving RE and ESSs [16] and the control problem of energy management in a microgrid [17]. Besides, multi-agent DRL is used to solve the partitioned voltage control problem in a large-scale DN [18]. Due to the inherent randomness and susceptibility to security risks associated with RL, the field of safe RL has emerged as a critical area of research in recent years [19], which can be divided into three types. The first type is to introduce additional security constraint penalties in the reward function [20], [21]. The second type is to use constrained Markov decision process (MDP) to form constrained RL [22], [23]. However, the above two methods only have a certain degree of safety in the later stages of the training process and cannot completely guarantee the safety of the action. The third type is to introduce a safety layer into RL, mapping actions to a safe space to ensure the safety of scheduling commands [24], [25]. This method can ensure the security of the training process, but it has a large computational burden because it uses quadratic programming for security projection.

### C. Contribution and Summary

In summary, optimization-based dispatch methods of ADN have strong safety assurance capabilities, but they are highly dependent on environmental models. RL has high decision-making efficiency and can adapt well to the inaccuracy of model parameters, though it falls short in terms of safety. Designing an approach to synergistically combine the strengths of the above two methods is the core idea of this study.

To achieve fast, safe and economical dispatch of ESSs in the ADN, this study proposes a complementary RL and optimization method. Compared to previous research, the primary contributions can be delineated as follows:
  i) Compared with the RL-only approach [17], [18], this study proposed a novel approach of complementary decision-making between DRL method and optimization method. The proposed approach effectively integrates the benefits of the DRL method for rapid decision-making, handling model inaccuracies and non-convexity, with the strengths of the optimization method in guaranteeing security.
  ii) Compared with the existing method of combining optimization and learning algorithms [26], [27], this study introduces a fast system security assessment method for automatic switching between RL and optimization methods. The proposed approach performs well in multiple scenarios such as the early stage of the training process, general scenarios, and unexperienced scenarios.
  iii) The proposed method is first applied to the coordinated dispatch problem of ESSs in ADN. The simulation result shows that this approach can achieve real-time, safe and economical dispatch.

The remainder of the article is structured as follows: Section II provides an overview of the essential background knowledge. In Section III, the methodology underlying the proposed approach is detailed. Section IV describes the procedures for training and implementing the proposed algorithm. Finally, Sections V and VI present case studies and conclusions, respectively.

## II. PRELIMINARIES

This section primarily presents the preparatory knowledge. A mathematical model for the sequential coordination dispatch problem of ESSs in the ADN is introduced in subsection *A*. Subsection *B* subsequently provides an exposition on DRL and MDP, and reframes the scheduling problem within the MDP framework.

### A. ADN Dispatch Problem

ADNs represent cutting-edge power distribution systems designed to facilitate bidirectional power flow and optimize the integration of RE sources. Employing advanced control methodologies and smart grid technologies, these systems aim to significantly enhance the reliability, efficiency, and sustainability of the power grid. In this study, the ADN mainly consists of grid structure, fluctuating loads, wind turbines (WTs), photovoltaics (PVs) and ESSs.

*1) Objective Function of the ADN dispatch problem*

The purpose of the ADN dispatching in this study is to use the ESSs to perform peak and valley arbitrage on the premise of ensuring safety, so as to minimize the overall cost [28]. The dispatch instruction execution interval is 1 hour. The objective formulation is shown as follow:

$$\min_{P_{ESS}} F = \min \sum_{t=0}^{T} \left[ C_r P_r(t) + \sum_{k=1}^{N_{ESS}} C_e(t) P_{ESS}(k,t) \right] \quad (1)$$

where $F$ denotes the operating cost of the ADN; $C_r$ indicates the electricity price of the grid connection node; $P_r(t)$ denotes the active power emitted by the grid connection node at time $t$; $C_e(t)$ denotes the electricity price of the node where the ESS located; $P_{ESS}(k,t)$ represents the output power of the $k^{th}$ ESS; $T$ is the overall dispatch period; $N_{ESS}$ is the number of ESSs.

*2) Modeling the operation environment of the ADN*

As for the ESSs in the ADN, the SoE of the ESS at the current moment is determined by the output power and duration of the previous moment. This change process can be expressed using the ampere-hour integral method [29]:

$$SoE(t+1) = \begin{cases} SoE(t) + \dfrac{1}{E_{ESS}} \cdot \eta^{ch} \cdot \int_t^{t+1} P_{ESS}(t)dt, P_{ESS}(t) \geq 0 \\ SoE(t) + \dfrac{1}{E_{ESS}} \cdot \dfrac{1}{\eta^{dis}} \cdot \int_t^{t+1} P_{ESS}(t)dt, P_{ESS}(t) < 0 \end{cases} \quad (2)$$

where $SoE(t)$ denotes the state of the energy of the ESS; $E_{ESS}$ denotes the energy capacity of ESS. $\eta^{ch}$ and $\eta^{dis}$ represent the charging and discharging efficiencies.

In Equation (2), the positive value of $P_{ESS}(t)$ indicates that the ESS is in a charging state, otherwise it is in a discharging state.

To guarantee the safe operation and longevity of ESSs, it usually has SoE safety range limits and output power safety range limits, which can be articulated as:

$$SoE_{\min} \leq SoE_t \leq SoE_{\max} \quad (3)$$

$$\underline{P}_{ESS}(t) \leq P_{ESS}(t) \leq \overline{P}_{ESS}(t) \quad (4)$$

$$\underline{P}_{ESS}(t) = \max(-P_{ESS}^{\max}, [SoE_{\min} - SoE(t)] \cdot E_{ESS} \cdot \eta_t^{ch}) \quad (5)$$

$$\overline{P}_{ESS}(t) = \min(P_{ESS}^{\max}, [SoE_{\max} - SoE(t)] \cdot E_{ESS} / \eta_t^{ch}) \quad (6)$$

where $SoE_{\min}$ and $SoE_{\max}$ denote the lower and upper bounds of SoE; $\underline{P}_{ESS}(t)$ and $\overline{P}_{ESS}(t)$ denote the lower and upper bounds of the ESS output power at time $t$; $P_{ESS}^{\max}$ indicates the maximum output power determined by the characteristics of the battery itself.

The power flow functions are formulated as follows:

$$V_{Re}(i,t)\sum_{j=1}^{N}[G_{ij}V_{Re}(j,t) - B_{ij}V_{Im}(j,t)] + \\ V_{Im}(i,t)\sum_{j=1}^{N}[G_{ij}V_{Im}(j,t) + B_{ij}V_{Re}(j,t)] + P(i,t) = 0, i \in N \quad (7)$$

$$P(i,t) = P_L(i,t) - P_{PV}(m,t) - P_{WT}(w,t) + P_{ESS}(k,t), \\ i \in N, m \in N_{PV}, w \in N_{WT}, k \in N_{ESS} \quad (8)$$

$$V_{Im}(i,t)\sum_{j=1}^{N}[G_{ij}V_{Re}(j,t) - B_{ij}V_{Im}(j,t)] - \\ V_{Re}(i,t)\sum_{j=1}^{N}[G_{ij}V_{Im}(j,t) + B_{ij}V_{Re}(j,t)] + Q(i,t) = 0, i \in N \quad (9)$$

$$Q(i,t) = Q_L(i,t), i \in N \quad (10)$$

$$\underline{V} \leq V(i,t) \leq \overline{V} \quad (11)$$

where $V_{Re}(i,t)$ denotes the real part of the complex voltage of bus $i$ at time $t$; $G_{ij}$ denotes the conductance of line $ij$; $B_{ij}$ denotes the susceptance; $V_{Im}$ is the imaginary part of the voltage; $P$ and $Q$ represent the injection values of the active and reactive power, respectively; $P_L$ and $Q_L$ denote the active and reactive power of the load demand; $P_{PV}(m,t)$ denotes the output power of the $m^{th}$ PVs; $P_{WT}(w,t)$ denotes the output power of the $w^{th}$ WTs; $N$ denotes the number of nodes in the ADN; $N_{PV}$ denotes the number of PVs; and $N_{WT}$ represents the number of WTs; $V$ indicates the voltage; $\underline{V}$ and $\overline{V}$ denote the lower and upper bounds of the voltage, respectively.

Equations (7) and (9) represent the active and reactive power flow functions. Equations (8) and (10) correspond to the injection values of active and reactive power. Equation (11) delineates the safety constraint for node voltage.

### B. DRL and MDP

Constrained by the energy limitations of ESSs, the issue addressed in this study exemplifies a classical sequential decision problem. RL is a kind of effective way to solve sequential problems [30]. By integrating the robust approximation capabilities of DNNs with the decision-making faculties of RL, DRL enables computers to achieve human-level intelligence across a variety of intricate tasks [31].

The RL takes an action by processing observations derived from the environmental state, with the objective of maximizing the cumulative reward. The cumulative reward represents an aggregate metric generated by the environment in reaction to the agent's executed actions.

The above process can be expressed as MDP. It can be presented as a tuple $\langle S, A, \mathbb{P}, r, \gamma \rangle$. Among them, $S$ and $A$ represent state and action space; $\mathbb{P}(\cdot|s,a): S \times A \to S'$ represents the state transition probability; $r$ denote the reward and $\gamma \in (0,1]$ represents the discount factor. The coordinated dispatch problem of ESSs in the ADN is reformulated as an MDP so that it can be solved by DRL.

State space: The observation is the useful information that can be obtained by the agent from the environment state. In this problem, we assume that the observation and state are equal. The state $s_t \in S$ is delineated as follow:

$$s_t = [\{P_L(i,t)\}_{i \in N}, \{P_r(t)\}, \{P_{ESS}(k,t)\}_{k \in N_{ESS}}, \{SoE(k,t)\}_{k \in N_{ESS}}, \\ \{\underline{P}_{ESS}(k,t)\}_{k \in N_{ESS}}, \{\overline{P}_{ESS}(k,t)\}_{k \in N_{ESS}}, \{C_e(T)\}_{T=t,t+1,\ldots,t+23}] \quad (12)$$

where $\{C_e(T)\}_{T=t,t+1,\ldots,t+23}$ represents a set comprising the present electricity price and the forecasted for the subsequent 23 hours.

Action space: The action $a_t \in A$ is consisted of the output power of the ESSs, which is shown as follow:

$$a_t = [\{P_{ESS}(k,t)\}_{k \in N_{ESS}}] \quad (13)$$

Reward: The reward $r_t$ denotes the environmental feedback after the agent executes action $a_t$ when the environment is in state $s_t$, which is defined as follows:

$$r_t = -(\frac{C_{total}}{C_W} + \delta) \quad (14)$$

$$C_{total} = \sum_{t=0}^{T}\left[C_r P_r(t) + \sum_{k=1}^{N_{ESS}} C_e(t) P_{ESS}(k,t)\right] \quad (15)$$

where $C_{total}$ is the total costs of the ADN operation in time range $T$; $C_W$ is the scaling factor for the total operating costs; $\delta$ is a penalty term [32], quantified by the frequency with which the voltages of the nodes exceed the specified limit at the current time.

State transition: The state transition probability $\mathbb{P}$ denotes the likelihood of transitioning from the current state $s_t$ to the subsequent state $s_{t+1}$ following the execution of action $a_t$.

The state of $P_L$ and $C_e$ are updated by the data set, and the $SoE$ is controlled by the $P_{ESS}$, which is shown as (2). $\underline{P}_{ESS}$ and $\overline{P}_{ESS}$ are determined by (5) and (6), respectively.

## III. METHODOLOGY OF THE PROPOSED COMPLEMENTARY REINFORCEMENT LEARNING AND OPTIMIZATION APPROACH

This section begins by presenting the framework of the proposed methodology for complementary RL and optimization approach in subsection *A*. Subsequently, it provides a detailed exposition of the three principal components in subsection *B*, *C* and *D*, which constitute the proposed approach.

### A. Framework Formulation

The framework of the proposed complementary RL and optimization approach to realize the coordinated dispatch strategy of ESSs in the ADN is shown in Fig. 1. This algorithm fully combines the real-time scheduling capabilities of DRL and the security assurance capabilities of the optimization method. It introduces the security assessment DNN as a means of judging whether the DRL dispatch commands are safe, thereby determining whether the optimization method needs to be used for secondary solution.

Initially, the trained DRL component produces a preliminary dispatch command based on the current environmental status, which may include potential security threats. Subsequently, both the current environmental status and the scheduling command generated by the DRL component are forwarded to the security assessment component. This component evaluates the safety of the scheduling command. If the safety criteria are satisfied, the command is deemed secure and executed. Conversely, if the command is considered unsafe, it is classified as a hazardous command, prompting the invocation of the optimization component to produce a secure scheduling command.

### B. Security Assessment Component

The security assessment component is used to promptly evaluate the security of the dispatch commands output by the DRL.

The conventional alternating current power flow (ACPF) calculation utilizing Newton's iteration method can precisely determine the voltage values at each node, enabling the identification of potential safety risks associated with limit violations. Nonetheless, this approach is computationally intensive and cannot satisfy the rapidity demands of real-time scheduling. Additionally, given that not all nodes are high-risk, it is often unnecessary to compute the voltage values for every node.

In view of this, this study adopts a security assessment component based on DNN. By incorporating the current state of the ADN (encompassing the active and reactive loads of each node and the renewable energy generation) along with the dispatch commands generated by the DRL component into the DNN, the voltage value of high-risk nodes can be determined. This process allows for an assessment of whether the commands from the DRL component meet safety standards. If the safety criteria are satisfied, the DRL commands will be executed directly. Otherwise, the optimization component will be activated to provide a feasible command for the current moment. The determination of whether a node is classified as high-risk is conducted through a combination of expert knowledge and data-driven methods.

The objective of training the DNN is to minimize the value of the loss function. This function is defined as the root mean square error (RMSE) between the actual voltage values of several high-risk nodes, as determined by the ACPF, and the output values produced by the DNN, as illustrated in (16):

$$Loss = \sqrt{\frac{1}{N_{risk}} \sum_{i=1}^{N_{risk}} (V_{i,pred} - V_{i,real})^2} \qquad (16)$$

where $N_{risk}$ is the number of the high-risk nodes; $V_{i,pred}$ denotes the voltage of the $i^{th}$ high-risk node obtained by the security assessment neural network; $V_{i,real}$ denotes the real voltage of the $i^{th}$ high-risk node calculated by the ACPF.

By using the security assessment neural network to replace the ACPF, the training time can be significantly shortened.

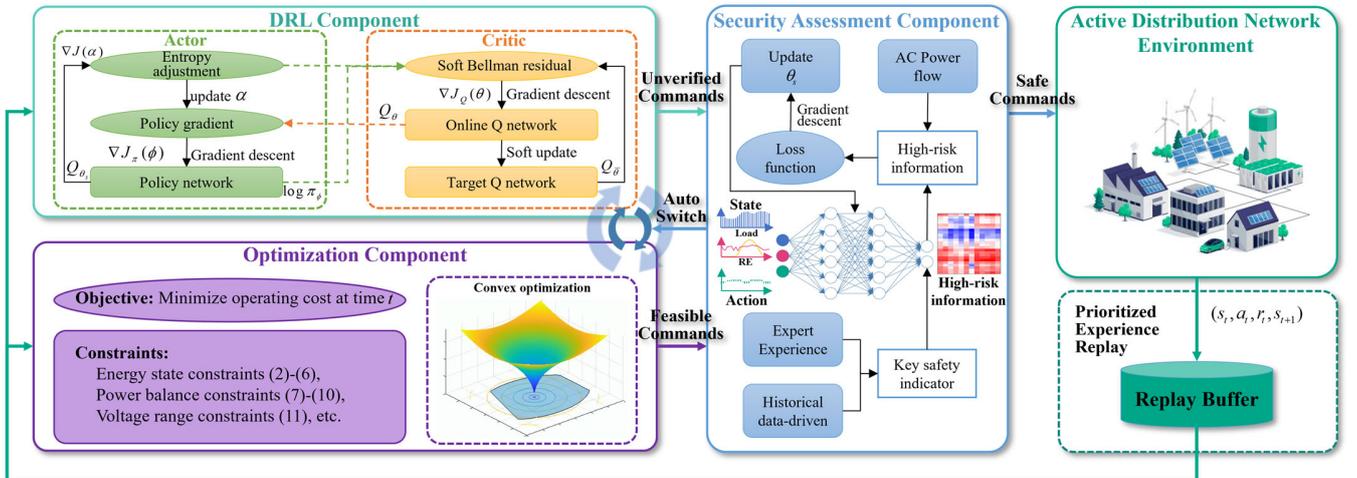

**Fig. 1.** The framework of the proposed complementary reinforcement learning and optimization approach.

## C. DRL Component

The Soft Actor-Critic (SAC) is one of the most efficient DRL algorithms with high training efficiency and promising application scenarios[33], [34], [35]. In light of this, the DRL component in this study adopts a SAC algorithm [36] improved by priority experience replay (PER) technology [31]. The objective function of SAC is to obtain an optimal policy that satisfies:

$$\pi^* = \arg\max_{\pi} \mathbb{E}_{(s_t,a_t)\sim\rho_\pi} \left[ \sum_{t=0}^{\infty} \gamma^t \left( r(s_t,a_t) + \alpha \mathrm{X}(\pi(\cdot|s_t)) \right) \right] \quad (17)$$

where $\pi^*$ represents the optimal policy; $\rho_\pi$ denotes the joint probability distribution; $r(s_t,a_t)$ denotes the reward accrued by the agent subsequent to executing action $a_t$ while the environment is in state $s_t$; $\alpha$ is the temperature coefficient, which is used to influence the weight of entropy; X denotes the entropy of the policy $\pi$ at state $s_t$.

The training process of the SAC can be expressed as two parts: policy evaluation and policy improvement.

*i) Policy evaluation part:* The soft Q-function parameters can be trained to minimize the soft Bellman residual as follows:

$$\min J_Q(\beta) = \mathbb{E}_{(s_t,a_t)\sim\mathbb{D}} \left[ \frac{1}{2} \left( Q_\beta(s_t,a_t) - Q_{soft}(s_t,a_t) \right)^2 \right] \quad (18)$$

$$Q_{soft}(s_t,a_t) = r(s_t,a_t) + \gamma \mathbb{E}_{S_{t+1}\sim\mathbb{P}} \left[ V_{soft}(s_{t+1}) \right] \quad (19)$$

$$V_{soft}(s_t) = \mathbb{E}_{a_t\sim\pi} \left[ Q_{soft}(s_t,a_t) - \alpha \log \pi(a_t|s_t) \right] \quad (20)$$

where $\mathbb{D}$ denotes the replay buffer. The iterative gradient update for the soft Q-network is expressed as follow:

$$\nabla_\beta J_{Qsoft}(\beta) = \nabla_\beta Q_\beta(s_t,a_t) \big( Q_\beta(s_t,a_t) - r(s_t,a_t) \\ - \gamma \big( Q_{\beta'}(s_{t+1},a_{t+1}) - \alpha \log \big( \pi_\omega(a_{t+1}|s_{t+1}) \big) \big) \big) \quad (21)$$

where $\beta$ and $\beta'$ indicate the parameters of the soft Q-network and the target Q-network; $\omega$ indicates the parameter of the policy network.

*ii) Policy improvement part:* The parameters of policy network can be trained to minimize the Kullback-Leibler divergence as follows:

$$\min J_\pi(\omega) = \mathbb{E}_{s_t\sim\mathbb{D}} \left[ D_{KL}\left( \pi_\omega(\cdot|s_t) \left\| \frac{e^{Q_\beta(s_t,\cdot)}}{Z_\beta(s_t)} \right. \right) \right] \quad (22)$$

where $Z_\beta(s_t)$ is partition function that used to normalize the distribution, it has no impact on the policy gradient descent update and can be ignored. During each iteration, the new policy $\pi_{new}$ is updated according to the following formula:

$$\pi_{new} = \arg\min_{\pi'\in\Pi} D_{KL}\left( \pi'(\cdot|s_t) \left\| \frac{e^{Q^{\pi_{old}}(s_t,\cdot)}}{Z^{\pi_{old}}(s_t)} \right. \right) \quad (23)$$

The policy network undergoes gradient updates iteratively, with each iteration characterized by the following procedure:

$$\nabla_\omega J_\pi(\varphi) = \nabla_\omega \log \pi_\omega(a_t|s_t) + \big( \nabla_{a_t} \log \pi_\omega(a_t|s_t) \\ - \nabla_{a_t} Q(s_t,a_t) \big) \nabla_\omega f_\omega(\varepsilon_t;s_t) \quad (24)$$

where $\varepsilon_t$ is an input random noise; $f_\omega(\varepsilon_t;s_t)$ is used to reparametrize the policy by:

$$a_t = f_\omega(\varepsilon_t;s_t) \quad (25)$$

By employing PER, RL algorithms can more efficiently leverage limited empirical data, expedite the learning process, and potentially enhance the quality of the final learned policy [31].

## D. Optimization Component

The objective of this component is to promptly formulate a viable dispatch command that adheres to security constraints in cases where the security assessment component identifies the scheduling command from the DRL as unsafe. Consequently, this component focuses solely on generating a command that satisfies immediate security requirements, without the necessity of optimizing across the entire scheduling cycle. The objective of this optimization component is expressed as follow:

$$\min_{P_{ESS}} F'(t) = \min \left[ C_r P_r(t) + \sum_{k=1}^{N_{ESS}} C_e(t) P_{ESS}(k,t) \right] \quad (26)$$

and the constraints are shown in (2)-(11). Due to the non-convex nature of power flow constraints, the Distflow model is employed for second-order conic convex relaxation [37].

## IV. TRAINING AND EXECUTION PROCESSES OF THE SA2CO ALGORITHM

The training process of the SA2CO algorithm is shown in Algorithm 1. In the initial phases of training, both the DRL component and the security assessment component are undergoing training concurrently. The security assessment component is used to output the voltage values of these high-risk nodes, and compared with the ACPF results to obtain the loss of the neural network. Then pass the loss backward to train the neural network. When the average loss is less than 1e-2, the security assessment neural network is considered to have completed training, and the network parameters will be saved. In the subsequent DRL component training process, the trained security assessment component is directly used to determine the security of the instructions.

---

**Algorithm 1**: training procedure of the proposed SA2CO

**Input**: $\beta$, $\beta'$, $\omega$, $\gamma$, $\tau$, $\varepsilon$, $\alpha$, $E$, $T$, learning rate
**Output**: $\pi$
1: Initialize the hyperparameters of neural network;
2: Initialize the ADN environment state;
3: **Initialize the replay buffer**: $\mathbb{D}$ and the mini-batch size;
4: **for** episode = 1: $E$ **do**
5:     Observe the initial state ($s_0$) of the ADN environment
6:     **for** $t = 1: T$ **do**
7:         DRL component take action $a_t$ on the basis of (17)
8:         Calculate the network loss of security assessment component by (16)
9:         **if** network loss < 1e-2 **do**
10:           Save the security assessment neural network model, generate high-risk information $V_i$
11:         **else do**
12:           Backpropagate the network loss to train the

|  |  |
|---|---|
| | security assessment neural network |
| 13: | Generate high-risk information by ACPF |
| 14: | **if** (11) is satisfied **do** |
| 15: | Execute the action $a_t$ generated by DRL |
| 16: | **else do** |
| 17: | Use optimization component to generate safe dispatch instructions |
| 18: | Execute the dispatch commands in the ADN |
| 19: | Obtain the reward $r_t$ and the next state $s_{t+1}$ |
| 20: | Store the state transition in replay buffer $\mathbb{D}$ |
| 21: | **end for** |
| 22: | Update the parameters of policy network by (18)-(21) |
| 23: | Update the parameters of soft Q-network by (22)-(24) |
| 24: | **end for** |

During the training process of the DRL component, the DRL attempts to take actions (scheduling instructions for the ESSs) based on the observed ADN status. The security assessment component is then used to evaluate the security of the actions taken by the DRL. If the security is satisfied, the dispatch instruction generated by the DRL will be executed directly; otherwise, the optimization component will be used to generate a dispatch command that is feasible at the current moment to ensure the safe and reliable operation of the ADN. Afterwards, the reward computed according to equation (14) is subsequently provided as feedback to the DRL component and stored within the replay buffer. This stored reward is utilized to facilitate the ongoing training and parameter adjustment processes of both the policy network and the soft Q-network within the DRL framework. When the DRL component reaches the maximum number of iterations, the training process is completed and the network parameters will be saved.

The execution flow chart of the SA2CO algorithm is illustrated in Fig. 2. Firstly, the DRL component takes actions based on the ADN status. Secondly, the security assessment component evaluates the security of the dispatch commands. If the security is met, the commands will be directly executed. Otherwise, use the optimization component to generate a feasible solution at the current moment and execute it as a dispatch instruction. In order to ensure the real-time dispatch capability, the optimization component is only used to generate a feasible solution and does not guarantee the global optimality of the dispatch instruction.

## V. CASE STUDY

The performance of the proposed approach is tested with numerical simulations in this section. The simulation environment is introduced in subsection *A*. The effectiveness, superiority, and scalability of the proposed approach are verified in subsections *B*, *C*, and *D*, respectively.

### A. Simulation Environment Settings

The proposed SA2CO algorithm is tested in the modified IEEE33-bw DN model, and its network topology is shown in Fig. 3. Several distributed RE sources and ESSs are configured in this ADN, including 3 WTs, 3 PVs and 4 ESSs. Its specific installation nodes and capacity configuration information are shown in TABLE I.

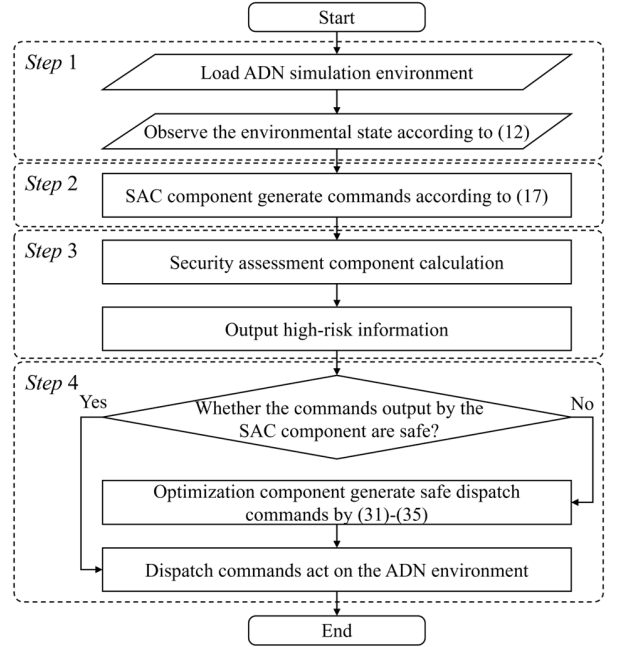

**Fig. 2**. The execution flow chart of the proposed SA2CO algorithm.

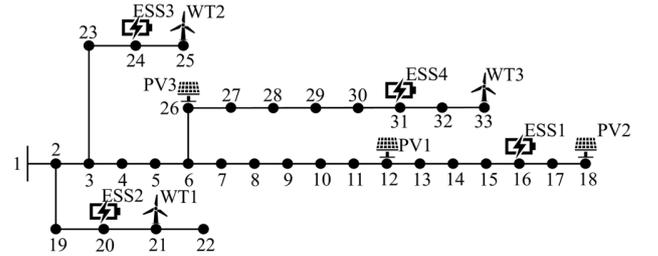

**Fig. 3.** The modified IEEE-33bw ADN topology diagram.

The initial SoE of all ESSs are set to 0.1, and the charging and discharging efficiencies ($\eta^{ch}$ and $\eta^{dis}$) are equal to 0.95. In addition, the $SoE_{min}$ for all ESSs are set to 0.1 and the $SoE_{max}$ are set to 0.9.

In this case, the electricity price data, RE data and load fluctuation data are all reasonably scaled values using the historical data (use the data from September 21 to October 10, 2019 as the training set) of the British electricity market, and the data can be accessed from the Open Power System Data platform.

TABLE I.
THE INSTALLATION LOCATION AND PARAMETERS OF DISTRIBUTED RESOURCES

| Devices | Node | Parameters | Value |
|---|---|---|---|
| WT1 | 21 | Maximum Power | 200kW |
| WT2 | 25 | Maximum Power | 400kW |
| WT3 | 33 | Maximum Power | 200kW |
| PV1 | 12 | Maximum Power | 100kW |
| PV2 | 18 | Maximum Power | 200kW |
| PV3 | 26 | Maximum Power | 200kW |
| ESS1 | 16 | Power/Capacity | 150kW/600kWh |
| ESS2 | 20 | Power/Capacity | 100kW/400kWh |
| ESS3 | 24 | Power/Capacity | 100kW/400kWh |
| ESS4 | 31 | Power/Capacity | 200kW/600kWh |

The security assessment neural network uses 66 neurons as the input layer (contains P and Q of all nodes). Combining historical data and expert experience analysis, nodes 12, 13, 14, 15, 16, 17, 18, 29, 30, 31, 32, and 33 in the 33-node system are high-risk nodes that are prone to voltage exceeding the limit [38]. Therefore, the output layer of the DNN contains 12 neurons, which are used to output the voltage information of 12 high-risk nodes. The learning rate (LR) is set to 2e-4, and the optimizer used in this case is Adam.

The hyperparameter settings of DRL component are shown in TABLE II. As shown in Fig. 1, the actor network within the DRL component receives as input the state of the ADN environment, which includes the load of all nodes at the current moment, the output power of all distributed generators at the current moment and the electricity price information for the next 24 hours. The output of the actor network is the distribution function of the dispatch commands for all ESSs and is activated using tanh units.

The input of the critic network in the DRL component is the state of the ADN environment and the action taken by the actor network in the current state. The output of the critic network is the value of $Q(s_t, a_t)$.

Both the actor network and critic network contain two hidden layers, using ReLU units as activation functions and Adam as optimizers.

TABLE II.
THE HYPERPARAMETER SETTINGS OF DRL COMPONENT

| Parameters | Value | Parameters | Value |
|---|---|---|---|
| Episode | 300 | Discount factor $\gamma$ | 0.99 |
| Steps in each episode | 480 | LR of Actor Network | 2e-4 |
| Layer size | 512 | LR of Critic Network | 2e-4 |
| Buffer size | 2e5 | Soft update factor $\tau$ | 1e-2 |
| Batch size | 64 | Weight decay factor | 1e-2 |

All the simulations are conducted on a computer with Intel i5-13500H and NVIDIA GeForce RTX 4060 to train the proposed SA2CO algorithm. The algorithm is programmed in Python, PyPower is used for alternating current power flow simulation, Pytorch is used to formulate the neural network, Gurobi solver is used to formulate and solve the optimization components.

### B. Effectiveness Analysis

The progression of the average cumulative reward throughout the training process is depicted in Fig. 4. The solid blue line indicates the average cumulative reward, while the blue shaded region denotes the standard deviation across five different random seeds. The average duration of the entire training process is 111.256 minutes.

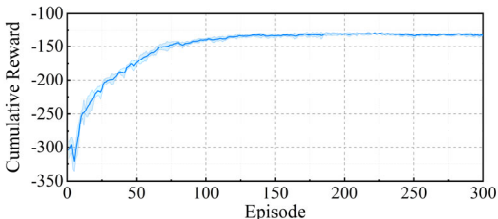

**Fig. 4.** The reward throughout the training process.

The proposed SA2CO algorithm used data from the UK electricity market from April 13 to August 30, 2020 as a test set. The test outcomes of the proposed algorithm across four representative scenarios are illustrated in Fig. 5. The simulation results in Figs. 5(a)-(d) are corresponding to the Cases #1-4, respectively. The upper part of each subgraph shows the fluctuation data of active load, PV and WT generation for 24 hours. Among them, the blue bar represents the active load fluctuation, the red line denotes the WT generation power, and the orange curve denotes the photovoltaic generation power. The lower part of each subgraph shows the 24-hour electricity price information of the day and the SoE of 4 ESSs. The red dotted line represents the electricity price, and the four bar of different colors represent the SoE of the four ESSs respectively.

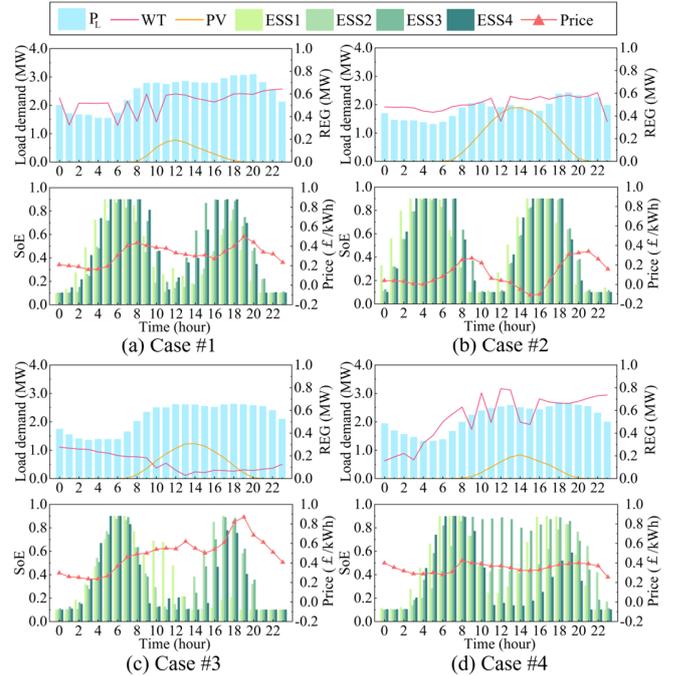

**Fig. 5.** Test results of the SA2CO approach in four typical scenarios.

Case #1 (Normal scenario): The proposed SA2CO algorithm demonstrates outstanding peak-valley arbitrage performance in this context. The four ESSs are in a charging state during the two low electricity price periods (2-5 a.m., 13-17 p.m.); during the two peak electricity price periods (7-10 a.m., 18-21 p.m.) are in a discharge state.

Case #2 (The large-scale photovoltaic power generation causes the electricity price to be negative at some times): The scenario where the electricity price is negative during part of the period has never appeared in the training set, the proposed algorithm still shows good arbitrage performance in this scenario.

Case #3 (The wind power shortage scenario): It can be seen from the Fig. 5(c) that from 7 to 12 am, the load demand increases rapidly, but the power generation of the WT decreases rapidly. Therefore, ESS1-ESS3 adjacent to the wind power are rapidly discharged from 7 to 12 am to maintain voltage stability. Due to the huge peak-to-valley price difference between 15 and 21 pm, this algorithm encourages ESSs to charge between 15 and 17 pm (balancing load demand by increasing the power purchased at the grid connection node of the DN) and discharge between 18 and 21 pm.

Case #4 (The electricity price is relatively stable): Since the power generation fluctuations of distributed power sources match the fluctuations of load demand, the electricity price fluctuates less at this time, and ESSs do not perform frequent charging and discharging operations.

The proposed algorithm was tested on the entire test set, and the resulting voltage distribution boxplot of 12 high-risk nodes is shown in the Fig. 6. As illustrated in the figure, the SA2CO algorithm demonstrates its safety by ensuring that the 12 high-risk nodes, which are susceptible to voltage exceedance, remain within the voltage limits throughout the hole test set.

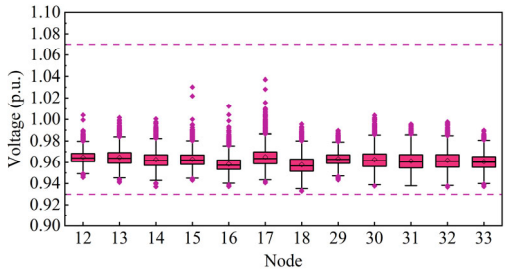

**Fig. 6.** Voltage distributions of 12 high-risk nodes in the distribution network.

### C. Superiority Analysis

To validate the efficacy of the proposed SA2CO algorithm, it has been benchmarked against several leading algorithms documented in the current literature. The comparison results are shown in the TABLE III.

Among them, the "Uncontrolled" means that ESSs in the ADN are always not active. In this case, the ADN can only purchase power from the root node and use the operating cost of one day as a benchmark value. In "Optimization", it is assumed that the accurate electricity price and RE power generation in the next day can be predicted to use the optimization method to schedule the ESSs. The operating cost is an ideal value (cannot meet the actual situation). The following four algorithms are real-time scheduling algorithms (only the electricity price prediction values of the next day, the load fluctuation value and the RE power generation are entered in real-time state). "SAC" is an algorithm that does not conduct a security assessment, and its reward is only related to operating costs; "ACPF-SAC" represents an algorithm that uses ACPF as a security assessment; "CSAC" also uses ACPF as a security assessment, the difference is that it uses constraint Markov decision process (CMDP) [39] to model, so it belongs to a constraint RL. "SA2CO" is an algorithm for the proposed complementary reinforcement learning and optimization approach in this study.

According to the comparison results, the proposed SA2CO algorithm offers better performance in terms of commands security, training and execution time. The use of the SA2CO algorithm for scheduling control can effectively reduce operating costs. Compared with the ideal optimal value of "Optimization", the effect of the proposed SA2CO algorithm in reducing operating costs is only 13.46% different. Compared with the optimization method, SA2CO does not rely on the predicted value of load and RE generation, which can achieve real-time scheduling. Compared with ACPF-SAC and CSAC, its overall training time is shortened by 22.08% and 25.99%, respectively.

TABLE III.
COMPARATIVE RESULTS OF DIFFERENT ALGORITHMS

| Algorithm | Training time(min) | Execution time(s) | Average total cost(£/day) | Improvement (%) |
|---|---|---|---|---|
| Uncontrolled | \ | \ | 1926.176 | \ |
| Optimization | \ | 52.4 | 1400.341 | 27.30 |
| SAC | 108.244 | 0.072 | 1634.761 | 15.13 |
| ACPF-SAC | 142.782 | 0.243 | 1638.218 | 14.95 |
| CSAC | 150.321 | 0.225 | 1684.205 | 12.56 |
| SA2CO | 111.256 | 0.115 | 1659.604 | 13.84 |

The number of unsafe dispatch commands in the training process of the four real-time scheduling algorithms is shown in the Fig. 7. SAC algorithm is easy to produce security risks in the whole training process, and the other three algorithms have a very low probability of security risks in the latter part of training, but it is not zero. The proposed SA2CO algorithm can quickly identify the security risks of DRL actions and switch to the optimization component to generate safe commands. In the early stage of algorithm training process, there are fewer unsafe dispatch commands because the dispatch potential reserved by ESSs is weak and the optimization method may not have a solution. Except this, the proposed approach can guarantee absolute safety in the middle and late stages of the training process.

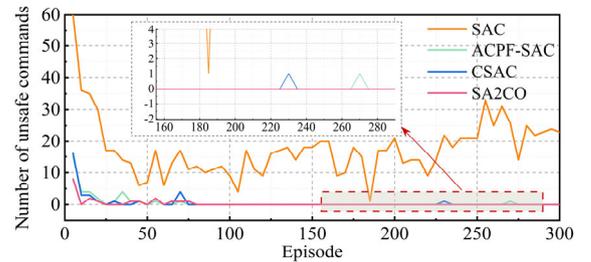

**Fig. 7.** The number of unsafe dispatch commands during the training process of algorithms.

### D. Scalability Analysis

To validate the scalability of the proposed method, an empirical evaluation is conducted in a 118-node distribution network system [40]. The ADN system contains 9 ESSs, 5 WTs and 5 PVs. The network topology and devices' allocations are presented in the Fig. 8.

The scheduling results of the proposed algorithm in the ADN system are shown in the Fig. 9. The load and RE generation fluctuation data used in this test case are the similar as those in Fig. 5(a). It can be seen from the test results that most ESSs charge in the low period of electricity price and discharge in the peak period of electricity price, which indicates that the proposed algorithm also has good scheduling ability in large-scale ADN systems.

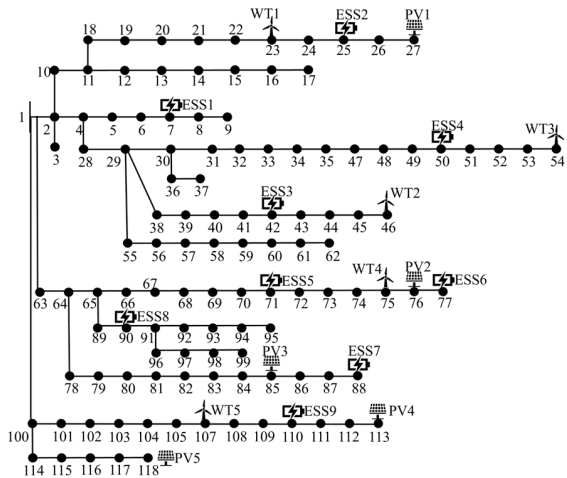

**Fig. 8.** Topology diagram of the 118-node ADN.

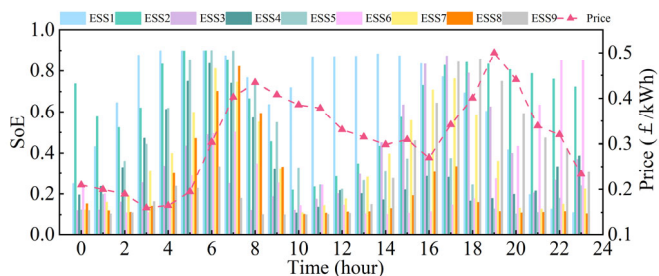

**Fig. 9.** Test results of the proposed algorithm in the 118-node ADN system.

## VI. Conclusion

This paper proposes a complementary decision-making framework that integrates optimization and RL. The security improvement architecture uses expert experience and data-driven training security assessment neural network embedded in the structure of DRL, which can ensure the fulfillment of real-time dispatch and security requirements. The proposed algorithm is applied to the coordinated dispatch problem of ESSs in the ADN, and compared with other representative optimization methods and RL-based methods in the state-of-the-art literature. Numerical simulation results indicate that the proposed method demonstrates superior efficiency in generating dispatch commands compared to the optimization method. Additionally, when contrasted with the RL-based approaches, the proposed method ensures the security of dispatch commands during the initial phases of algorithm training and within multifaceted complex scenarios.

## References

[1] G. He, J. Lin, F. Sifuentes, X. Liu, N. Abhyankar, and A. Phadke, "Rapid cost decrease of renewables and storage accelerates the decarbonization of China's power system," *Nat. Commun.*, vol. 11, no. 1, p. 2486, May 2020.

[2] A. Ehsan and Q. Yang, "State-of-the-art techniques for modelling of uncertainties in active distribution network planning: A review," *Applied Energy*, vol. 239, pp. 1509–1523, Apr. 2019.

[3] J. Cao, D. Harrold, Z. Fan, T. Morstyn, D. Healey, and K. Li, "Deep Reinforcement Learning-Based Energy Storage Arbitrage With Accurate Lithium-Ion Battery Degradation Model," *IEEE Trans. Smart Grid*, vol. 11, no. 5, pp. 4513–4521, Sep. 2020.

[4] A. Ehsan, Q. Yang, and M. Cheng, "A scenario-based robust investment planning model for multi-type distributed generation under uncertainties," *IET Generation, Transmission & Distribution*, vol. 12, no. 20, pp. 4426–4434, Nov. 2018.

[5] M. H. Lipu, S. Ansari, M. S. Miah and K. Hasan, "A review of controllers and optimizations based scheduling operation for battery energy storage system towards decarbonization in microgrid: Challenges and future directions," *Journal of Cleaner Production*, vol. 360, p. 132188, Aug. 2022.

[6] Y. Li, Y. Gu, G. He, and Q. Chen, "Optimal Dispatch of Battery Energy Storage in Distribution Network Considering Electrothermal-Aging Coupling," *IEEE Trans. Smart Grid*, vol. 14, no. 5, pp. 3744–3758, Sep. 2023.

[7] W. S. Ho, S. Macchietto, J. S. Lim, H. Hashim, Z. Ab. Muis, and W. H. Liu, "Optimal scheduling of energy storage for renewable energy distributed energy generation system," *Renewable and Sustainable Energy Reviews*, vol. 58, pp. 1100–1107, May 2016.

[8] L. Urbanucci, "Limits and potentials of Mixed Integer Linear Programming methods for optimization of polygeneration energy systems," *Energy Procedia*, vol. 148, pp. 1199–1205, Aug. 2018.

[9] J.-H. Teng, S.-W. Luan, D.-J. Lee, and Y.-Q. Huang, "Optimal Charging/Discharging Scheduling of Battery Storage Systems for Distribution Systems Interconnected with Sizeable PV Generation Systems," *IEEE Trans. Power Syst.*, vol. 28, no. 2, pp. 1425–1433, May 2013.

[10] T. Logenthiran, D. Srinivasan and T. Shun, "Demand Side Management in Smart Grid Using Heuristic Optimization," *IEEE Trans. Smart Grid*, vol. 3, no. 3, pp. 1244-1252, Sep. 2012.

[11] J. Hu, Y. Shan, J. M. Guerrero, A. Ioinovici, K. W. Chan, and J. Rodriguez, "Model predictive control of microgrids – An overview," *Renewable and Sustainable Energy Reviews*, vol. 136, p. 110422, Feb. 2021.

[12] I. L. R. Gomes, R. Melicio, and V. M. F. Mendes, "A novel microgrid support management system based on stochastic mixed-integer linear programming," *Energy*, vol. 223, p. 120030, May 2021.

[13] J. F. Franco, L. F. Ochoa, and R. Romero, "AC OPF for Smart Distribution Networks: An Efficient and Robust Quadratic Approach," *IEEE Trans. Smart Grid*, vol. 9, no. 5, pp. 4613–4623, Sep. 2018.

[14] E. DallAnese, K. Baker, and T. Summers, "Chance-Constrained AC Optimal Power Flow for Distribution Systems with Renewables," *IEEE Trans. Power Syst.*, vol. 32, no. 5, pp. 3427–3438, Sep. 2017.

[15] J. Schulman, F. Wolski, P. Dhariwal, A. Radford, and O. Klimov, "Proximal Policy Optimization Algorithms.", 2017, *arXiv: 1707.06347*.

[16] D. Cao, W. Hu, X. Xu, Q. Wu and Q. Huang, "Deep Reinforcement Learning Based Approach for Optimal Power Flow of Distribution Networks Embedded with Renewable Energy and Storage Devices," *Journal of Modern Power Systems and Clean Energy*, vol. 9, no. 5, pp. 1101–1110, Sep. 2021.

[17] S. Lee, J. Seon, Y. G. Sun, S. H. Kim and C. Kyeong, "Novel Architecture of Energy Management Systems Based on Deep Reinforcement Learning in Microgrid," *IEEE Trans. Smart Grid*, vol. 15, no. 2, pp. 1646–1658, Mar. 2024.

[18] D. Cao, J. Zhao, W. Hu, F. Ding and Q. Huang, "Data-Driven Multi-Agent Deep Reinforcement Learning for Distribution System Decentralized Voltage Control with High Penetration of PVs," *IEEE Trans. Smart Grid*, vol. 12, no. 5, pp. 4137–4150, Sep. 2021.

[19] L. Brunke, M. Greeff, A. W. Hall and Z. Yuan, "Safe Learning in Robotics: From Learning-Based Control to Safe Reinforcement Learning," *Annual Review of Control Robotics and Autonomous Systems*, vol. 5, pp. 411–444, May 2022.

[20] W. Wang, N. Yu, Y. Gao, and J. Shi, "Safe Off-Policy Deep Reinforcement Learning Algorithm for Volt-VAR Control in Power Distribution Systems," *IEEE Trans. Smart Grid*, vol. 11, no. 4, pp. 3008–3018, Jul. 2020.

[21] S. Lee and D.-H. Choi, "Federated Reinforcement Learning for Energy Management of Multiple Smart Homes with Distributed Energy Resources," *IEEE Trans. Ind. Inf.*, vol. 18, no. 1, pp. 488–497, Jan. 2022.

[22] H. Zhang, X. Sun, M. H. Lee, and J. Moon, "Deep Reinforcement Learning-Based Active Network Management and Emergency Load-Shedding Control for Power Systems," *IEEE Trans. Smart Grid*, vol. 15, no. 2, pp. 1423–1437, Mar. 2024.

[23] J. Zhang, L. Sang, Y. Xu, and H. Sun, "Networked Multiagent-Based Safe Reinforcement Learning for Low-Carbon Demand Management in Distribution Networks," *IEEE Trans. Sustain. Energy*, pp. 1–16, Jan. 2024.

[24] Z. Yi, X. Wang, C. Yang, C. Yang, M. Niu, and W. Yin, "Real-Time Sequential Security-Constrained Optimal Power Flow: A Hybrid Knowledge-Data-Driven Reinforcement Learning Approach," *IEEE Trans. Power Syst.*, vol. 39, no. 1, pp. 1664–1680, Jan. 2024.
[25] Y. Gao and N. Yu, "Model-augmented safe reinforcement learning for Volt-VAR control in power distribution networks," *Applied Energy*, vol. 313, p. 118762, May 2022.
[26] K. Li and J. Malik, "Learning to Optimize." ,2016, *arXiv: 1606.01885*.
[27] L. Sang, Y. Xu, Z. Yi, L. Yang, H. Long, and H. Sun, "Conservative Sparse Neural Network Embedded Frequency-Constrained Unit Commitment with Distributed Energy Resources," *IEEE Trans. Sustain. Energy*, vol. 14, no. 4, pp. 2351–2363, Oct. 2023.
[28] D. Metz and J. T. Saraiva, "Use of battery storage systems for price arbitrage operations in the 15- and 60-min German intraday markets," *Electric Power Systems Research*, vol. 160, pp. 27–36, Jul. 2018.
[29] T. Morstyn, B. Hredzak, R. P. Aguilera, and V. G. Agelidis, "Model Predictive Control for Distributed Microgrid Battery Energy Storage Systems," *IEEE Trans. Contr. Syst. Technol.*, vol. 26, no. 3, pp. 1107–1114, May 2018.
[30] Q. Fu, Z. Han, J. Chen, Y. Lu, H. Wu, and Y. Wang, "Applications of reinforcement learning for building energy efficiency control: A review," *Journal of Building Engineering*, vol. 50, Jun. 2022.
[31] V. Mnih, K. Kavukcuoglu and D. Silver, "Human-level control through deep reinforcement learning," *Nature*, vol. 518, no. 7540, pp. 529–533, Feb. 2015.
[32] Z. Ren, D. Dong, H. Li, and C. Chen, "Self-Paced Prioritized Curriculum Learning with Coverage Penalty in Deep Reinforcement Learning," *IEEE Trans. Neural Netw. Learning Syst.*, vol. 29, no. 6, pp. 2216–2226, Jun. 2018.
[33] F. Mao, Z. Li, Y. Lin, and L. Li, "Mastering Arterial Traffic Signal Control with Multi-Agent Attention-Based Soft Actor-Critic Model," *IEEE Trans. Intell. Transport. Syst.*, vol. 24, no. 3, pp. 3129–3144, Mar. 2023.
[34] Z. Zhang, Z. Chen, and W.-J. Lee, "Soft Actor–Critic Algorithm Featured Residential Demand Response Strategic Bidding for Load Aggregators," *IEEE Trans. on Ind. Applicat.*, vol. 58, no. 4, pp. 4298–4308, Jul. 2022.
[35] S. Wang, R. Diao, C. Xu, D. Shi, and Z. Wang, "On Multi-Event Co-Calibration of Dynamic Model Parameters Using Soft Actor-Critic," *IEEE Trans. Power Syst.*, vol. 36, no. 1, pp. 521–524, Jan. 2021.
[36] T. Haarnoja, A. Zhou, P. Abbeel, and S. Levine, "Off-Policy Maximum Entropy Deep Reinforcement Learning with a Stochastic Actor", 2018, *arXiv: 1801.01290*.
[37] H. Gao, J. Liu, L. Wang, and Z. Wei, "Decentralized Energy Management for Networked Microgrids in Future Distribution Systems," *IEEE Trans. Power Syst.*, vol. 33, no. 4, pp. 3599–3610, Jul. 2018.
[38] Y. Chai, L. Guo, C. Wang, Z. Zhao, X. Du, and J. Pan, "Network Partition and Voltage Coordination Control for Distribution Networks with High Penetration of Distributed PV Units," *IEEE Trans. Power Syst.*, vol. 33, no. 3, pp. 3396–3407, May 2018.
[39] C. Tessler, D. J. Mankowitz, and S. Mannor, "Reward Constrained Policy Optimization.", 2018, *arXiv: 1805.11074*.
[40] L. Sang, Y. Xu, and H. Sun, "Encoding Carbon Emission Flow in Energy Management: A Compact Constraint Learning Approach," *IEEE Trans. Sustain. Energy*, vol. 15, no. 1, pp. 123–135, Jan. 2024.